\documentstyle{l-aa}    
\def\ltsim{\raise 2pt \hbox {$<$} \kern-1.1em \lower 4pt \hbox {$\sim$}}
\def\ltapprox{\raise 2pt \hbox {$<$} \kern-1.1em \lower 5pt \hbox {$\approx$}}
\def\gtsim{\raise 2pt \hbox {$>$} \kern-1.1em \lower 4pt \hbox {$\sim$}}
\def\gtapprox{\raise 2pt \hbox {$>$} \kern-1.1em \lower 5pt \hbox {$\approx$}}
\def\arcsec{$^{\prime\prime}$}
\def\arcmin{$^{\prime}$}

\def\degrees{$^{\circ}$}
\def\skuno{\vskip 20pt}

\def\p0{\phantom{0}}

\def\ph1{\phantom{1}}
\begin{document}
\thesaurus{11.09.1 3C~449;13.18.1; 11.10.1; 11.13.2}
\title
{VLA Observations of the giant Radio Galaxy 3C~449}
\skuno
\skuno
\author{L. Feretti\inst{1}  \and R. Perley\inst{2} 
\and G. Giovannini\inst{1,3} 
\and H. Andernach\inst{4} }
\offprints{lferetti@ira.bo.cnr.it}
\institute{
Istituto di Radioastronomia del CNR, Via P. Gobetti 101, I-40129 Bologna, 
    Italy
\and
National Radio Astronomy Observatory, PO Box O,
Socorro, NM 87801 0387.
\and
Dipartimento di Fisica, Universit\'a di Bologna, 
I-40100 Bologna, Italy.
\and
Depto. de Astronom\'\i a, IFUG, Apartado Postal 144,
Guanajuato, C.P. 36000, Mexico. }

\maketitle

\begin{abstract}

In this paper we present multifrequency VLA observations of the
giant radio galaxy 3C~449. High resolution and sensitivity data were obtained
in total intensity and polarized flux at 1.365, 1.445, 1.465, 1.485, 4.685,
4.985 and 8.385 GHz. 

The source is characterized by an unresolved core, two opposite 
symmetric jets, and very extended lobes. 
These new images show the source morphology in great detail,
in particular we detect plumes and wiggles in the low brightness lobes.
The source is considerably polarized at all frequencies. We find
that the magnetic field orientation is parallel to the jet axis at the
very beginning of the jet, and  becomes perpendicular at about 10\arcsec
(5 kpc) from the core.
In the low brightness regions, the magnetic field is circumferential to the
edges of the emission region.

The spectral index map between 5 GHz and 8.4 GHz shows that the
northern region beyond 1\arcmin~ from the core 
 has a much flatter spectrum than the southern one.

We analyse the jet dynamics, and conclude that the jets are relativistic
at the beginning, and decelerate significantly within 10\arcsec(5 kpc)
 from the core.

We compute the rotation  measure, using all seven  frequencies between
1.4 and 8.4 GHz. We detect significant rotation measure with values
within about $\pm$50 rad m$^{-2}$ of the expected contribution of our
Galaxy. The
Faraday effect is likely to originate in an external screen, which is
most probably the intergalactic medium of the group of galaxies
around 3C~449.
Moreover, we resolve the effect of the interstellar medium associated
with the galaxy 3C~449 itself.

\keywords{Galaxies: individual: 3C~449 -- Radio continuum: galaxies --
Galaxies: jets -- Galaxies: magnetic fields} 

\end{abstract}

\section{Introduction}

The radio galaxies of luminosity lower than 
P$_{178}$=5$\times$10$^{25}$ W/Hz, known as FR\,I objects 
(Fanaroff \& Riley 1974)
are often characterized by continuous two-sided jets running into large-scale
lobe structures (plumes) which are edge-darkened (i.e with the brightness 
peak located less than halfway out from the nucleus) and whose steepest 
radio spectra lie in the outermost regions furthest from the host galaxy.
The magnetic field in FR\,I jets is oriented predominantly along 
the jet at the beginning, becoming predominantly perpendicular to the 
jet further away from the core (Bridle \& Perley 1984).

The observed morphologies of low-power jets have been used to argue that they
are  turbulent and propagating at sonic or  transonic speed
on the large scale (Bicknell 1984, Bicknell et al. 1990). 
On the other hand, recent observational work (Giovannini et al. 1995)  
leads to the scenario in which
jets of low-luminosity radio galaxies are relativistic on  parsec scales.
This is consistent with unified models, where FR\,I radio 
galaxies form the parent population
of BL Lac objects (Urry \& Padovani 1995).

Parma et al. (1994) have examined the variation of jet sidedness ratio
with distance from the core, total radio power and core prominence. Their
results are consistent with the 
hypothesis that FR\,I jets slow down from $\beta
\approx$ 0.6 to sub-relativistic velocities on scales of 1 -- 10 kiloparsecs,
because of turbulent entrainment.
Laing (1996) developed 
a two-component jet model consisting of a fast ``spine'' with
perpendicular magnetic field, surrounded by a slower  
``shear layer'', which interacts with the external medium. 
The ``shear layer'' has either parallel  or two-dimensional
magnetic field. Assuming that the
jet is relativistic at the beginning  and decelerates 
to non-relativistic speeds at large distances from the nucleus,
net longitudinal or transverse apparent fields may result, as a 
consequence of relativistic aberration. 

\begin{table*}
\caption{VLA Observing Log}
\begin{flushleft}
\begin{tabular}{lcclrccccc}
\hline
\noalign{\smallskip}
 Frequencies & Band & A  & A   & B  & B  & C   & C 
  & D  & D  \\
 (MHz) &  & Date & Time(h) & Date & Time(h) & Date & Time(h) & Date 
& Time(h) \\
\noalign{\smallskip}
\hline
\noalign{\smallskip}
 1365/1445 & L & APR94 & 1.5 & APR93 & 5.1 & JUL93 & 1.0 & 
  JAN94 & 0.6 \\
 1465/1485 & L & APR94 & 1.6 & APR93 & 5.1 & JUL93 & 1.0 &
  JAN94 & 0.5 \\
 4685/4985 & C & MAR/APR94 & 0.8 & APR93 & 5 & JUL93 & 8.7 &
    JAN94 & 1.8 \\
 8285/8485 & X & MAR94 & 1 & APR93 & 5.1 & JUL93 & 5.5 &
   JAN94 & 1.8 \\
\noalign{\smallskip}
\hline
\noalign{\smallskip}
\end{tabular}
\end{flushleft}
\end{table*}

The symmetric twin-relativistic jet model is supported by several lines
of evidence (see e.g. Laing et al. 1996), in particular the detection
of superluminal motions. However, 
it is well established that some sources show components with
subluminal motions, as it is the case e.g. for Centaurus A 
(Tingay et al. 1998), M~87 (Biretta  1996),
Cyg A (Carilli et al. 1994).
The subluminal jet components can be either interpreted as slow patterns
on the relativistic flow, or due to non-relativistic nuclear
ejection. The last possibility has been suggested by Sol et 
al. (1989), Pelletier \& Roland (1989, 1990) and Pelletier \& Sol (1992),
who proposed the jet two-fluid model to explain the presence 
of compact structures moving at different speed in extragalactic jets.

A typical radio galaxy of the FR\,I class is 3C~449,
which is well known due to its very large angular size
($\sim$ 30\arcmin). It is elongated
in the NS direction and is characterized by long two-sided jets.
It is relatively nearby
(z=.0181), and therefore particularly suitable for
a detailed study of the jet structure and of the polarization
properties.
Previous studies of this source at radio frequencies have been
presented by  many authors (see Andernach et al. 1992, and references
therein). High resolution images were first presented by Perley et al. (1979), 
who studied the jet structure.
The high symmetry of the 3C~449 jets has been taken as evidence for
the radio source major axis lying close to the plane of the sky. 
The parent galaxy, UGC12064, is dumb-bell and is
the most prominent member of the Zwicky cluster 2231.2+3732.
It shows a nuclear dust lane with a major axis diameter of $\sim$2\arcsec~ 
roughly in 
position angle 135\degrees~ (Butcher et al. 1980). The optical isophotes
show pronounced deviations from pure ellipses and gradients in ellipticity,
which are clear indications of a strong gravitational
interaction (De Juan et al. 1994, Balcells et al. 1995).
Optical data obtained with the HST were presented by Capetti et
al. (1994), who detected a 23 mag nucleus surrounded by a ring
with  a projected diameter
of 0.4\arcsec. The luminosity profile shows that the observed ring is actually
the result of absorption. The size of this region is consistent with
it being cold material
associated with an extended accretion disk.

In this paper we present multifrequency VLA maps, at high resolution and
sensitivity, which allow us to study the jet behaviour and the
polarization.
We use a Hubble constant H$_0$=50 km s$^{-1}$ Mpc$^{-1}$, which implies
a linear conversion to 0.52 kpc/arcsec~ at the distance of 3C~449.

\section {Observations and data reduction}

The data presented here were obtained with the Very Large Array (VLA)
in all four configurations, at different frequencies in 
the L, C and X bands. The observing epochs and integration times are listed
 in Table 1. The source  3C286 was used as a primary flux
density calibrator. The phase calibrator was the nearby point source
2200+420, observed at intervals of about 20 minutes, 
while the calibrators for the polarization position angle were
 3C138 and 3C48. The on-axis instrumental polarization of the antennas
was corrected using the secondary calibrator 2200+420, which was
observed over  a wide range of parallactic angles.
The data were reduced with the Astronomical Image Processing System
(AIPS), following the standard procedure: Fourier-Transform, Clean and
Restore. Self-calibration was applied to minimise the effects of
amplitude and phase uncertainties.
The X-band data obtained at 8285 GHz and 8485 GHz were averaged.
The data-sets from different configurations  were 
first reduced separately, each
undergoing several iterations of cleaning and phase self-calibration.
At 8.4 GHz and 5 GHz, the unresolved radio nucleus was subtracted in each 
individual  data-set to avoid undesirable effects of core flux 
variability in the final maps. 
Table 2 lists the core flux densities subtracted at each frequency
and epoch. To produce the merged data at each frequency, 
the D and C data-sets were concatenated first, then the B
data-set, and finally the A data-set were added. At each step the data-sets
were self-calibrated to ensure phase consistency.

\begin{table}
\caption{Core flux densities}
\begin{flushleft}
\begin{tabular}{llll}
\hline 
\noalign{\smallskip}
 Date   & Conf. & S$_{5 GHz}$ & S$_{8.4 GHz}$ \\ 
        &       &  mJy        &  mJy   \\
\noalign{\smallskip}
\hline
\noalign{\smallskip}
 APR93   & B &     31.0  &  40.7  \\
 JUL93   & C &     37.0  &  45.0  \\
 JAN94   & D &     36.0  &  45.0  \\
 MAR94   & A &     37.6  &  43.4   \\
 APR94   & A &     38.7  &  ~~--    \\
\noalign{\smallskip}
\hline
\noalign{\smallskip}
\end{tabular}
\end{flushleft}
\end{table}

At each of the seven frequencies, maps
of the Stokes parameters I,
Q and U were produced  with three different resolutions,
using the AIPS task IMAGR. The restoring beam was a circularly
symmetrical Gaussian, with FWHM = 1.25\arcsec, 2.5\arcsec~ and 5\arcsec.
The images with the highest resolution at 8.4 GHz 
(0.5\arcsec~ and 0.8\arcsec) and 4.9 GHz (0.8\arcsec) were produced 
by applying the Maximum Entropy Method (task VTESS), to 
ensure the proper imaging of resolved low-brightness  structure.
The images of the polarized intensity were obtained as
$ P =  (Q^2 + U^2)^{1/2}$, and corrected for the positive Ricean bias
due to this combination of two noisy quantities in quadrature
(Wardle \& Kronberg 1974).  The polarization angle was derived
according to $\theta = 0.5 \tan ^{-1} (U/Q)$.

\section {Results}

\subsection {Total intensity images}

The overall source structure is easily visible at 1.4 GHz, at a
resolution of 5\arcsec~ (Fig. 1).
The source is characterized by an unresolved core, two opposite symmetric jets,
extending for $\sim$ 1\arcmin~ from the core, and the lobes of
lower brightness. The total extent of the 
source mapped at this frequency with the VLA is  about 
20\arcmin, which is slightly smaller than the extent 
found from single dish images
(Andernach et al. 1992). In Fig. 1 some relevant source regions 
are labelled for further reference. 
The two innermost jets (N1 and S1)
seem fairly symmetrical at this relatively low resolution, although they are
not perfectly opposite to each other. They are misaligned by 
$\sim$11\degrees, as mentioned by Perley et al. (1979). 
They both merge at about the same distance from the core 
into prominent inner lobes of high brightness  (N2 and S2), 
different in structure.  
On larger scales,  beyond the inner lobes N2 and S2, the source
is rather asymmetrical. In the northern
part the structure widens (region N3), then  bends to the east
into region N4, at which point there is a sharp bend of 
90\degrees~ bend to NW. The eastern boundary 
of region N4 resembles the corner of a square.
Further out, the structure (N5) is first oriented in the NS direction, 
and then  bent in a far trail toward NE. This region
has a very low brightness, and smoothly fades into the noise.
In the southern part, the structure emerging from the inner lobe S2 and
directed toward SW (region S3) seems to be a channel with edge
brightening, and becomes 
narrower at increasing distance from the core. 
Further to the south this streamer merges into a large,
almost round lobe of very low brightness (S4). 

The overall low-brightness structure
is characterized by wiggles, filaments and plumes. There is a  remarkable,
faint, arc-like feature,  westward of the southern lobe S2. 
It is actually detached from the lobe and it is very unusual.
An I-band CCD image of this region  was obtained
with the 88-inch telescope of the University of Hawaii (see Fig. 2), 
and no object clearly responsible for this emission has been found. 
Therefore, we believe this structure to be associated with 3C~449. 

In the field mapped at 1.4 GHz, there is an extended unrelated
source at  RA(B1950)=22$^h$ 28$^m$ 36.5$^s$, DEC(B1950)=
39\degrees~ 08\arcmin~ 47.7\arcsec~, which shows a
double-lobed structure, with a  total flux of 12.5 mJy. It has no optical
counterpart in the R plate of the Palomar Digitized Sky Survey, 
and must surely be an unrelated background radio galaxy.

\begin{figure}
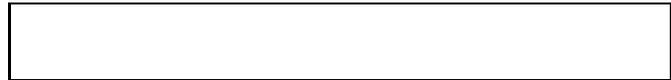

\picplace{1.0cm} 
\caption{Contour map of 3C~449 at 1.365 GHz with 5\arcsec~
angular resolution. Contour levels are 0.1, 0.25, 0.5, 1, 2.5, 5, 10,
25, 50, and 100 mJy/beam. The rms noise level in the map is 0.035 
mJy/beam.}
\end{figure}

\begin{figure}
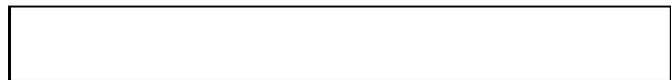

\picplace{1.0cm} 
\caption{1.4 GHz radio contours of the region of the arc-like feature
are superposed onto an I-band 15 min CCD image taken at the UH 88\arcsec~ 
telescope.
}
\end{figure}

\begin{figure}
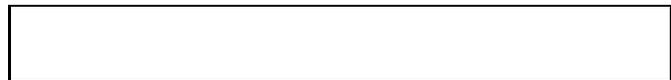

\picplace{1.0cm} 
\caption{Grey scale image at 1.365 GHz with 1.25\arcsec~ angular resolution.
The grey scale range is 0.05 to 2 mJy/beam. The rms noise level 
in the map is 0.031 mJy/beam.
}
\end{figure}

At higher resolution (Fig. 3), 3C~449 is detected only in the innermost 
8\arcmin,  with great structural details.
The jets show some level of asymmetry in  brightness, but they
are very similar in  transverse size. The two jets are not 
straight, but show a gradual  curvature to the east. This behaviour
is responsible for the  misalignment of the two jets detected
at lower resolution.
At approximately  $\sim$45\arcsec~ north and south of the core, the
two jets show a prominent  bend westward. 
After the bend, the northern jet widens and forms
the higher brightness inner lobe
(N2 in Fig. 1), which has a sharp and round edge on its NW boundary. 
From there the streamer continues due NE and narrows 
at the same time up to a well-defined leading edge, before it
sharply bends toward NW again. The whole structure is reminiscent
of a helix.
The southern jet at 45\arcsec~ from the core
has a sharper westward bend than the northern jet, 
and seems more collimated even after the bend. After a smooth bend
along a long arch towards E, it forms the
bright inner lobe (S2 in Fig. 1) which has a higher brightness than the
corresponding northern one. 
The jet can still be distinguished within the lobe, where it follows a 
curved trajectory. Also noticeable is a loop emerging from the
eastern side of the S2 lobe, suggesting a helical motion. 
The channel in the region S3 clearly exhibits enhanced emission at the 
edges.
The western edge seems to be the continuation of the north-eastern loop,
while the eastern one emerges from the lower brightness southern part
of the lobe. We are  possibly seeing two source regions in projection
along the line of sight.

The features detected at 1.4 GHz are easily  visible in the maps
at  high frequencies, where the nucleus has been subtracted because
of its small, but noticeable variability (see flux densities in Table 2).
Fig. 4 gives the image at 5 GHz, where the source
is detected over  a total extent of $\sim$8\arcmin.
The details are strikingly similar to those at lower frequencies,
with filaments and plumes. 

\begin{figure}
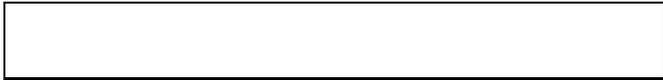

\picplace{1.0cm} 
\caption{Image of the source at 4.985 GHz with 2.5\arcsec~ 
angular resolution. The unresolved radio core
has been subtracted. Contour levels
are -0.05, 0.05, 0.1, 0.15, 0.3, 0.5, 0.75, 1, 2.5, and 3.5 mJy/beam.
The rms noise level in the map is 0.018 mJy/beam.
}
\end{figure}

\begin{figure}
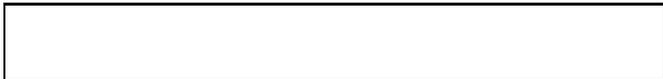

\picplace{1.0cm} 
\caption{Grey scale image of the source at 8.4 GHz with 0.8\arcsec~
resolution, overlayed onto the optical image from the red Palomar Digitized
Sky Survey. The unresolved radio core has
been subtracted. The grey scale range is 0.02 to 0.3 mJy/beam. 
The rms noise level in the map is 0.011 mJy/beam.  
}
\end{figure}

\begin{figure}
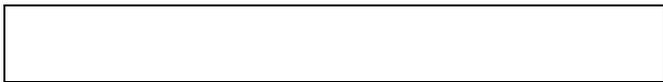

\picplace{1.0cm} 
\caption{Grey scale image of the innermost jets at 8.4 GHz, at the highest
available angular resolution of 0.5\arcsec. The unresolved core has
been subtracted at the position marked by a gapped X sign.
The grey scale range is 0.01 to 0.2 mJy/beam. 
The rms noise level in the map is 0.011 mJy/beam.
}
\end{figure}

The maps at highest resolution are those where the jets are heavily
resolved. Figs. 5 and 6 present the images at 8.4 GHz
with resolutions of 0.8\arcsec~ and 0.5\arcsec, respectively, 
obtained with the Maximum Entropy Method. 
These maps clearly show the structure of the jets, which are exactly 
aligned only within 10\arcsec-12\arcsec~ from the core, with a position
angle of $\sim$9\degrees.
The position angles of the northern and southern jet on a larger scale are
$\sim$13\degrees~ and $\sim$182\degrees, respectively. 
In Fig. 5, the radio image is overlaid onto the optical image
from the Digitized Palomar Sky Survey. The ring of absorbing material
(dust lane),  as found by Butcher et al. (1980), is not visible
here. It is easily visible in the image available in the public
archive of the Hubble
Space Telescope, and is located toward 
SW, roughly in position angle 135\degrees~ and 
is therefore not perpendicular to the jets, as found in other FR\,I 
radio galaxies (Capetti \& Celotti 1998). 

Some degree of asymmetry between the two jets is evident, with 
the southern jet being  slightly brighter on average than the
northern one. 
The two jets are of very low brightness  at their beginning, and
no gap of radio emission is found close to the nucleus. 
At $\sim$2\arcsec~ from the core, the southern
jet  shows a brightening, which 
has no similar counterpart in the opposite jet.
Both jets suddenly flare and widen, at about 8\arcsec~ from the core.
The brightness of the southern jet shows a distinct maximum at
$\sim$10\arcsec, followed first by a decrease in brightness, and then
by a transverse structure similar to a bow-shock at 
$\sim$35\arcsec~  from the core.
In the  northern jet, the trend of brightness is smoother.
The jets are easily distinguishable for about 1\arcmin, where they merge
into the inner lobes.

\subsection {Spectrum}

The spectrum of 3C~449 between 330, 1445,  and 4835 MHz, with
3.6\arcsec~ resolution, has been
analyzed in detail by Katz-Stone \& Rudnick (1997, hereinafter KR).
Here, we obtained the map of spectral index between 4.985 GHz and 8.4 GHz 
with 2.5\arcsec~ resolution (see Fig. 7). 

\begin{figure}
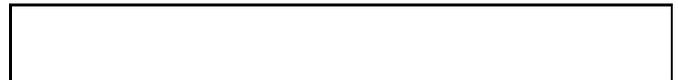

\picplace{1.0cm} 
\caption{Map of the spectral index between 4.985 GHz and 8.4 GHz, with
2.5\arcsec~ resolution. The core has been subtracted. The contour
is drawn at the level of 0.4.
}
\end{figure}

The spectrum is constant along the two  jets, including the 
bends at $\sim$45\arcsec~ from the core,  with an average spectral index
$\alpha^{8.4}_{5}$ = 0.58$\pm$0.03 ($S_{\nu}\propto \nu ^{-\alpha}$). 
In the northern inner lobe (N2), beyond the bend of the jet, the
spectrum index is still rather flat, with an average value of
 0.57$\pm$0.03, and becomes even flatter  at the NW 
boundary. In the southern inner lobe (S2), 
the spectrum is much steeper than in
the jet. In the center and southern edge of the lobe, where  
the jet is still traceable either from the high resolution  images and from the
polarization behaviour, the spectral index is 0.74$\pm$0.03, 
while on both sides  it increases up to $\sim$0.9.
There is therefore a remarkable asymmetry in the spectral behaviour
 between the northern and southern regions: the spectrum of the 
northern inner lobe is as
flat as that of the jet, while the southern one is characterized
by a significant steepening.

\subsection {Polarization}

The source is considerably  polarized at all frequencies. 
Fig. 8 presents the 8.4 GHz map, with the magnetic field vectors
superimposed (corrected for RM, see next subsection). 
The orientation of the magnetic field is 
longitudinal at the very beginning of the jet, where the polarization
percentage is lower, and becomes
transversal at larger distance from the core,  and in the bending regions
at $\sim$45\arcsec~ from the core.  
In the SW part of the lobe N2 the magnetic field is transversal to the
main ridge of emission, while in the NE part of the northern inner lobe
N2, after the bend of the structure by 90\degrees, the magnetic field 
is oriented along the main ridge of emission. 
In the southern inner lobe (S2) the magnetic field is transversal 
over most of the lobe, confirming that this is the 
continuation of the jet, but it is 
circumferential to the structure in the southernmost edge of lobe S2.

The polarization
percentage  at 8.4 GHz (Fig. 9) is somewhat patchy, and shows similar
behaviour in the northern and southern source regions.
 In the  innermost weak  jets within few arcsec from the
core, the degree of  polarization is around 15\% on both
sides. Further out, the jet polarization percentage increases to  about
30\%, then it drops to 15-20\% in the bends. In the inner lobes N2 and S2,
the values of the degree of polarization are  30-50\%. 
The map of the polarization percentage enhances the structures detected
in total intensity: in particular the bow structure in the southern jet 
($\sim$32\arcsec~ from the core) is 
prominent for its high fractional polarization and 
 a similar feature is now discernable in the northern jet. 
Also, the loop in the
southern inner lobe is very well  seen in the polarization percentage map
and a similar loop is found at the northern boundary of the northern inner
lobe. We note that the polarization percentage is 
fairly constant transverse to the jet. Only the NW edge of the 
inner lobe N2 shows enhancement.

The source is not significantly  depolarized at 4.985 and 4.685 GHz.
At the lower frequencies, around 1.4 GHz, the average
polarization percentage is about 0.75-0.8 of that at 8.4 GHz in the
jets, while in the inner lobes (N2 and S2) no significant
depolarization is found.
It is worth noting  that polarized brightness at the level of 30-40\% is
also detected in the large scale structure (Fig. 10), which is
not imaged at higher frequencies because of its low brightness and
missing short spacings. We also note that the two bright edges of the
channel in the region S3 (see Fig. 1) are well separated because of their 
high degree of polarization and a similar feature of two parallel
polarized streamers is found to emerge northward from the inner lobe N2.

\begin{figure}
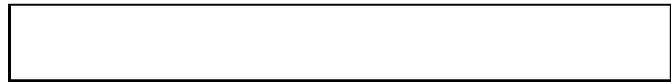

\picplace{1.0cm} 
\caption{Map at 8.4 GHz with 1.25\arcsec~ resolution, with 
vectors indicating the projected magnetic field direction, 
corrected for RM.  The rms noise level in the total intensity
map is 0.011 mJy/beam.
The horizontal bar at the center of the right panel marks the
position where the unresolved core has been subtracted. In both
images, contours levels are
-0.06, 0.06, 0.15, 0.3, 0.5, 0.75, and 1 mJy/beam. 
The vectors are proportional in
length to the polarization percentage, with 1\arcsec~ corresponding to 
20\% in the left panel, and 1\arcsec~ corresponding to 25\% in the right panel.
}
\end{figure}

\begin{figure}
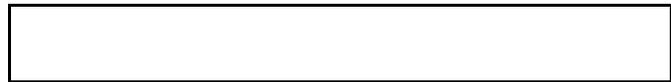

\picplace{1.0cm} 
\caption{Grey scale image of the fractional polarization at 8.4 GHz with 
1.25\arcsec~ resolution.  The unresolved nucleus has been subtracted.
The grey scale levels is 0 to 50\%.
}
\end{figure}

\begin{figure}
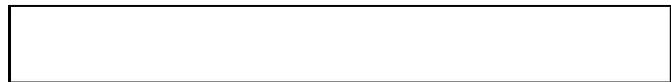

\picplace{1.0cm} 
\caption{The 1.485 GHz map, with 2.5\arcsec~ resolution, showing 
the polarization vectors (E field, uncorrected for RM). They are 
proportional in length
to the fractional polarization, with 1\arcsec~ corresponding to 5\%.
Contour levels are -0.1, 0.1, 0.5, 1, 5, and 10 mJy/beam.
The rms noise level in the total intensity map is 0.032 mJy/beam.
}
\end{figure}

\subsection {Rotation measure}

We  obtained an image of the rotation measure (RM),
 by combining the maps of the polarization {\bf E} vector at the seven
frequencies available to us. 

We used the maps with all available resolutions (1.25\arcsec,
2.5\arcsec, and 5\arcsec), after blanking the pixels where the
uncertainty in the polarization angle exceeded 20\degrees.
Following the definition  $\theta_{\lambda} = RM \lambda^2 $, where
$\theta_{\lambda}$ is the position angle of the polarization vector at
the wavelength $\lambda$, the value of the RM at each valid pixel 
was computed by linear fitting of 
the polarization  angle as a function of $\lambda^2$.
The fitting algorithm provides a weighted least-squares fit, allowing
for an ambiguity of $\pm$n$\pi$ in each polarization position angle.

\begin{figure}
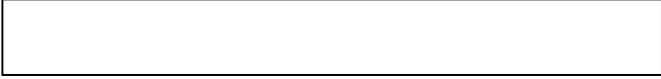

\picplace{1.0cm} 
\caption{Samples of the fits to the polarization angle, to obtain the
rotation measures. These values refer to the region of the southern jet,
at $\sim$8\arcsec~ from the nucleus.
}
\end{figure}

\begin{figure}
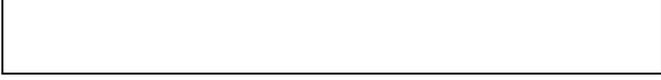

\picplace{1.0cm} 
\caption{Histogram of the rotation measure for all significant pixels
in the source
}
\end{figure}

\begin{figure}
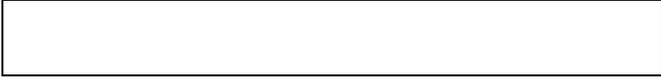

\picplace{1.0cm} 
\caption{Map of the rotation measure at 1.25\arcsec~ resolution,
computed using all 7 frequencies
ranging from 1.365 GHz to 8.4 GHz. 
}
\end{figure}

The data are well fitted by a $\lambda^2$ relation, as shown by some sample
fits plotted in Fig. 11. The formal errors on the
best-fit results are typically less than 2 rad m$^{-2}$.
The values of RM range between -210 and -100 rad m$^{-2}$ (see
histogram in Fig. 12), with a peak around -162 rad
m$^{-2}$, which corresponds to the expected foreground RM of our
Galaxy (Andernach et al. 1992). The distribution of the RM is
non-gaussian, with an excess of more negative values with respect to the
average. 
The map of RM with the highest angular resolution (1.25\arcsec)
is given in Fig. 13.
A remarkable feature in the RM map is the
large symmetry in the inner jets,  within $\sim$15\arcsec~ from the
core, where the mean RM is  more negative than the Galactic
value ($<$RM$>$ = --198 $\pm$ 3 rad m$^{-2}$).

At larger distance from the core,
the RM in both jets flips to values more positive than the Galactic value
and a clear  asymmetry between the two jets becomes
most evident in the bends.  In the northern jet, there is some
transversal structure and the values of RM change
from --146  $\pm$ 10 rad m$^{-2}$ to   --185 $\pm$ 6 rad m$^{-2}$
in the bending region. In the southern jet, the values of RM are
always more positive than the Galactic value, from  --152 $\pm$ 5 
to  --116 $\pm$ 5 immediately before the bending, and  --138 $\pm$ 5 
rad m$^{-2}$ in the  bending region.

The two inner lobes  N2 and S2 are quite similar, with  the average
RM close to the Galactic value. 
The maps of the RM obtained with lower resolution are in excellent
agreement with the map at 1.25\arcsec~ resolution, indicating that 
the fluctuations of RM are resolved. 
Variations of the RM occur on similar scales in the jets and
lobes, with typical size of \gtsim 10\arcsec, i.e about 5 kpc.

\section {Discussion}

\subsection {Jet collimation}

We used the 8.4 GHz maps with
0.5\arcsec~ resolution to derive brightness profiles perpendicular
to the jet axes. These were fitted with a Gaussian function to 
obtain the  FWHM $\Phi_{obs}$ and peak surface brightness I$_{obs}$ 
of the jets at various
distances from the radio core. In order to derive 
intrinsic quantities, these parameters 
were  deconvolved of the CLEAN beam $\Phi_{beam}$, with the
following formulae (Killeen et al. 1986):
$$ \Phi = (\Phi^2_{obs}- \Phi^2_{beam})^{1/2} \eqno(1)$$
$$ I_\nu = I_{obs}(1+\Phi^2_{beam} \Phi^{-2})^{1/2} \eqno(2)$$

These parameters are plotted in
Fig. 14, where the dots and open circles refer to the southern and
northern jet, respectively. The two jets show  very similar behaviours
in the collimation properties, which can be divided into three regimes:
an opening angle of $\sim$8.5\degrees~ at the jet beginning,
a  rapid expansion between 6\arcsec~(3 kpc) and 10\arcsec~(5 kpc)
from the core, with
an opening angle of $\sim$17\degrees, and finally a recollimation 
beyond 10\arcsec~, with an opening angle of $\sim$7.5\degrees.
The rapid expansion coincides with 
the region where the jet magnetic field changes its orientation, from
longitudinal to transversal (see Fig. 8).

The   brightness is different in the two jets, and is  
higher, on average,  in the southern jet. 
In both jets, the overall trend of brightness against 
width is very different from that expected in an adiabatic jet with constant
velocity (see Subsect. 4.4).

\begin{figure}
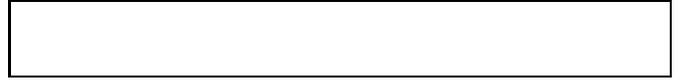

\picplace{1.0cm} 
\caption{Plot of the jet deconvolved FWHM versus distance (left), and
of the jet peak brightness versus FWHM (right), with 0.5\arcsec~
resolution. Dots and open circles refer
to the southern and northern jet, respectively. The dashed line in the left
panel represents the best fit to the data in the 3 collimation regimes.
}
\end{figure}

\subsection {Core prominence}

Assuming that the jets are relativistic at their origin, information
on their orientation with respect to the line of sight
can be inferred  from the comparison between the core radio
power and the total radio power, following the approach of 
Giovannini et al. (1994). This argument is based on the fact that
the core radio emission contains a Doppler-boosted
relativistic jet, whose strength depends on the jet inclination  to 
the line of sight.  Given the existence in radio galaxies
of a general correlation between the core power at 5 GHz, P$_c$, and the
total radio power at 408 MHz, P$_{tot}$  (Giovannini et al. 1988):
$$ LogP_c = 11.01 + 0.47 Log P_{tot} \eqno(3)$$
the expected intrinsic core power can be inferred from the total low frequency 
radio power, which is not affected by Doppler boosting. 
The P$_c$-P$_{tot}$ correlation given in Eq. (3) 
has been obtained using a large sample of radio galaxies
oriented at random angles. Therefore, it gives the apparent (beamed)
core radio power for a galaxy at the average orientation of 60$^{\circ}$.
The Doppler enhancement in a jet of velocity $\beta$c
and spectral index $\alpha$, oriented at an angle $\theta$
to the line of sight is
$$ P(\theta) = P[\Gamma (1 - \beta cos \theta)]^{-(2 + \alpha)} \eqno(4)$$
where $\Gamma$ is the Lorentz factor, $\Gamma$=(1-$\beta^2$)$^{-1/2}$. 
Therefore, the ratio of
the measured core power P$_{c-obs}$ to that inferred from Eq. (3),
P$_{c-exp}$, corresponding to $\theta$=60\degrees, is given by:  
$$ {P_{c-obs} \over P_{c-exp}} = \left({1 - \beta cos \theta \over 
1 -0.5 \beta }\right)^{-(2 + \alpha)} \eqno(5)$$
and  provides an estimate of  $\beta$ and $\theta$.

We derived the ratio of the measured to expected core power using 
the core flux densities at 5 GHz obtained from the present data, and the
total flux density at 408 MHz given in Andernach et al. (1992).
A statistical uncertainty of 1 r.m.s was taken into account
in the flux density measurements, but the most relevant parameter
in the determination of this ratio  was  the core
variability (see Tab. 2 for the values of core flux densities).
We obtained that P$_{c-obs}$/P$_{c-exp}$  is in the range 0.33-0.45.
The corresponding allowed region for the two parameters $\beta$ and $\theta$,
assuming $\alpha$=0, 
is presented in Fig. 15. It implies a large velocity of the jet at its
beginning, and an  angle  larger than 75\degrees,
between the  source major axis and the line of sight.

\begin{figure}
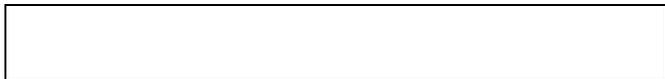

\picplace{1.0cm} 
\caption{Allowed region (shaded) for the jet velocity and inclination to the
line of sight, from the core prominence relative the total flux.
}
\end{figure}

\subsection {Jet sidedness}

Independent constraints on $\beta$ and $\theta$ can be
obtained from the jet to counter-jet ratio.
The relativistic beaming effect on  two relativistic
symmetrical jets,  is to enhance the 
surface brightness of the approaching beam, and decrease that of 
the receding one. The observed  jet to counter-jet ratio R is given by
the formula 
$$ R = (1+\beta cos \theta)^{q} (1-\beta cos \theta)^{-q} \eqno(6)$$
where $q$ = 2+$\alpha$ for an isotropic jet with no
preferred magnetic field direction (Pearson \& Zensus 1987), and 
$q$= 3+2$\alpha$ for a jet with a
perfectly ordered  longitudinal magnetic field (Begelman 1993).
Since the magnetic field is predominantly transversal, and the degree
of polarization is low in the region where it is longitudinal (see Sect. 3.3)
the formula valid for the isotropic jet is the most appropriate for the
present source.

Following Laing (1996), we produced an image of  ``jet-sidedness'', 
using the 8.4 GHz map at 0.5\arcsec~ resolution, 
and considering  the southern jet as the main jet (see Sects. 3.1 and 4.1). 
Since the two jets show a slightly
curved path (see e.g. Fig. 6), the sidedness image is only meaningful 
in their innermost region, where they are perfectly straight.
The image is given in Fig. 16.
The average of the values in this map is 1.4, with no distinct trend either
along or across the jet. The bright points
at the lateral edges could be affected by the slight curvature of the jets.

Fig. 17 gives the run of the jet to the counter-jet ratio
along the ridge of maximum brightness.
The low values close to the nucleus are in the region where the core 
has been subtracted, and should not be trusted.
This plot cannot be simply interpreted in the framework of the
relativistic beaming effects, since it seems to be stronlgy affected
by fluctuations related to local enhancements in the jet brightness.
The highest value of the jet to counter-jet ratio is $\sim$2.1, and
corresponds to the bright spot at about 2\arcsec~ from the core,
which could be a knot of emission. If instead it is due to 
relativistic beaming,
it implies a value of $\beta cos \theta$=0.14, which leads to a jet
velocity of $\approx$ c for a viewing angle of $\approx$ 82\degrees.
The moderate jet asymmetry derived from  Figs. 16 and 17
confirms that 
the source must be at a large angle to the line of sight.

\begin{figure}
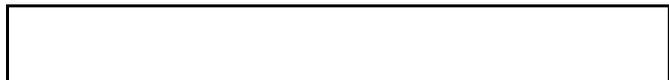

\picplace{1.0cm} 
\caption{Map of the jet to counter-jet ratio, computed at 8.4 GHz with 
0.5\arcsec~ resolution. The coordinate scale is labelled in arcsec relative
to the core position
}
\end{figure}

\subsection {An adiabatic model for the jets}

Information on the jet dynamics has been obtained following
the simple approach
that the jet is  expanding adiabatically, conserving the number of
relativistic particles and frozen in magnetic field. Under this assumption, 
the jet velocity, brightness and opening angle are
related. The functional dependence between these parameters has been
discussed in the limit of non-relativistic bulk motion 
 with  3-D expansion by Fanti et
al. (1982), Bicknell (1984) and Perley et al. (1984). The case of
relativistic bulk motion 
is considered by Baum et al. (1997), who
obtain the following relationships:

Predominantly parallel {\bf B} field: 
$$I_{\nu} \propto 
(\Gamma_j v_j )^{-(2 \alpha +3)/3} r_j^{-(10 \alpha
+9)/3} D^{2+\alpha}  \eqno(7)$$

Predominantly transverse {\bf B} field: 
$$I_{\nu} \propto 
(\Gamma_j v_j )^{-(5 \alpha +6)/3} r_j^{-(7 \alpha
+6)/3} D^{2+\alpha} \eqno(8)$$

\noindent
where $I_{\nu}$ is the jet surface brightness, 
$\alpha$ is the  spectral index, 
v$_j$ and $\Gamma_j$ are the jet velocity and Lorentz factor, and 
D is the Doppler factor $D=(\Gamma_j(1-\beta cos \theta))^{-1}$. 

According to these formulae, the widening of the jet at constant velocity
results in strong adiabatic losses, which would cause its dramatic dimming.
By deceleration of the advance speed, these adiabatic losses can be
balanced  and the jet brightness maintained. The apparent brightness of
the jet is also affected by relativistic Doppler boosting. 

Using the above relationships, we have modeled the jet intrinsic
brightness and width as a function of distance, to derive the jet velocity.
In this procedure, the input parameters are the orientation
of the magnetic field, taken from the from the observations,
the initial jet velocity, $\beta_i$, and the inclination angle of the 
jet with the respect to the line of sight, $\theta$,
which are both assumed.

The jet observational parameters  are well reproduced by the
trends given in Fig. 18 (see caption), which correspond to pairs of 
the initial velocity and inclination angle consistent with 
the constraints obtained by the core prominence argument.
We modeled both the (southern) jet, and the (northern) counter-jet,
using the appropriate orientation, and we obtained
very similar trends. The important result is 
that always the jet velocity strongly decreases  
with distance up to $\sim$10\arcsec~ (5 kpc) from the core,  
and shows almost constant values further out. 

The jet/counter-jet ratios  derived from the velocity trends displayed
in Fig. 18 are plotted  in Fig. 17, for a 
comparison with the observed values (continuous line). 
The two models with lower initial velocity are more appropriate
to reproduce the jet to counter-jet ratios.
They imply that the initial peak is
mostly due to a local knot, and also the jet asymmetry at 
10\arcsec-13\arcsec~ from the core cannot be ascribed to  Doppler
boosting. 

Assuming the orientation of 3C~449 of $\sim$82.5\degrees~
to the line of sight, and the jet velocity running from 0.9c
to 0.7c in the inner 5\arcsec-6\arcsec~ (Fig. 18), the Doppler
factor  is in the range 0.5-0.9. A Doppler factor
lower than 1 means that the jets are strongly de-boosted and the
observed brightness is lower than their intrinsic brightness. 
This is in agreement with the fact that the innermost jets show very low
brightness, and could
explain the presence of a ``gap'' region between the nucleus and the jets
in many FR\,I sources.

\begin{figure}
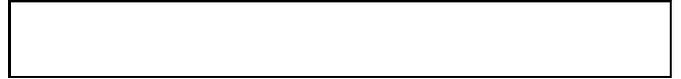

\picplace{1.0cm} 
\caption{Brightness ratio of the jet to counter-jet as a 
function of the distance from the core, along
the ridge of source maximum brightness (continuous line). 
Typical 1$\sigma$ errors are about 0.06 close to the nucleus,
and progressively decrease with distance down to 0.02.
The dashed lines correspond to the expected ratios for 
the jet models given in Fig. 18 (with the same meaning of the line-types). 
}
\end{figure}

\begin{figure}
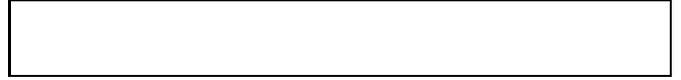

\picplace{1.0cm} 
\caption{Trend of $\beta$ as a function of the distance from the core, in the
adiabatic model. The different line-types refer to the following models
(consistent with the allowed region of Fig. 15): 
short-dash: $\theta$ = 85\degrees and $\beta_i$ = 0.85;
long-dash: $\theta$ = 82.5\degrees and $\beta_i$ = 0.90;
dot-dash: $\theta$ = 80\degrees and $\beta_i$ = 0.95.
}
\end{figure}

We finally note that acceptable fits to the observational jet 
parameters can  be also obtained
using  values of the initial jet velocity lower than those implied
by the core prominence. In particular, it is interesting that
starting with $\beta_i$=0.5, the jet velocity beyond 10\arcsec~ drops to 
$\beta \sim$0.01, which would be consistent with the
velocity inferred by Hardee et al. (1994) from the observed trajectory 
of the southern jet, as a result of the orbital motion of the central
engine. This case, however, does not account for the innermost low
brightness of the jets as due to Doppler de-boosting, and can only
be reconciled with the core prominence argument
assuming that a strong jet deceleration takes place
very close to the nucleus.

Our main conclusion is that the simple adiabatic model 
provides  evidence  for a jet deceleration within a few 
arcsec, i.e. a few kpc, from the nucleus. 
The strong jet deceleration occurs in the region where the magnetic field
changes the orientation, from longitudinal to transversal.
A likely mechanism for the jet deceleration is the
interaction between the jet and the external medium, i.e. the  
entrainment of material from the surrounding medium. 
In particular, the pressure of the gaseous atmosphere associated 
with the galaxy, would confine the jets and affect the jet
dynamics (Bicknell et al. 1990). 
This behaviour is in agreement with
the unified models, and seems to be typical in FR\,I radio jets, 
as modeled for 3C~31 by Laing (1996). In the Laing
model, the jet consists of a central spine, with higher velocity and
transverse magnetic field, and a shear-layer with lower velocity and
either a longitudinal or a two-dimensional field. We do not detect  
these two components in the inner jets, either in the 
sidedness image, or in the structure of the magnetic field. 
Hardcastle et al. (1997) found a similar problem in 3C~296.
In 3C~449, the large orientation angle of the source with respect to 
the line of sight could make it difficult to distinguish  the two 
components, since the Doppler boosting is not strong enough to significantly
enhance the high velocity component with respect to the low velocity one.

\subsection {Case of non-relativistic nuclear ejection}

The discussion in the above subsections was based on the assumption 
that the nucleus of 3C~449 ejects highly relativistic jets.
Since no VLBI observations are presently available for 3C~449, the
emission of the jets at non-relativistic speed 
in principle cannot  be ruled out.
An explanation of the structure of 3C~449, assuming that the nucleus
ejects a moderately relativstic electron-proton jet with v$_j \simeq$
0.4c, has been provided by Roland et al. (1992).
In their scenario, if the mass ejected is large enough and the jet is well
collimated, the jet kinetic energy density $\rho_j$v$^2_j$ 
(where $\rho_j$ is the jet 
density) is much greater than the external pressure P$_{ext}$.
This is the case for the innermost region, i.e. within 6\arcsec~ from the
core, where the jets have low brightness and small opening angle.
As the jets open, for a critical radius of the jets, the kinetic energy 
density becomes comparable to the pressure of the external medium.
Here, the jets interact with the outer medium and dissipate
their kinetic energy via turbulence. The relativistic electrons
are accelerated by the turbulence in the dissipation area, which
corresponds to the bright jets up to about 1\arcmin~ from the core. 
Beyond this distance,  the dissipation process is finished and
the jet velocity becomes smaller or 
equal to  the sound speed of the external medium.

\subsection {Spectral behaviour}

KR studied the spectral behaviour of 3C~449 over 3 frequencies (0.3, 
1.4, and 4.8 GHz), using a new analysis tool, the spectral tomography, 
which allowed them  to isolate the contribution of two different
spectral components, the ``flat jet'', and the ``sheath''. The 
flat jet remains fairly well
collimated  within 2.5\arcmin~ from the core,  is characterized by
a power-law spectrum with $\alpha$ = 0.53$\pm$0.01, and shows
little steepening with distance from the core. The second component,
the ``sheath'',  appears
beyond 1\arcmin~ from the core, on both sides of the jet, 
and is responsible for most of the observed widening.
The spectrum of the ``sheath'' is steeper than that of the
``flat jet'', and has a range of spectral indices
with typical values at 2.5\arcmin~ from the core of 0.57$\pm$0.02 in 
the north, and 0.69$\pm$0.02 in the south.
We refer the reader to the paper by KR for the interpretation of the two jet 
components. 

Our spectrum between 4.985 GHz and 8.4 GHz (Fig. 7)
reveals the existence of a spectral asymmetry
between the northern and southern inner lobes. A similar
asymmetry is found by KR in the sheath.
For the flat jet, i.e. within 1\arcmin~ from the core, we obtain 
$\alpha^{8.4}_{5}$
= 0.58$\pm$0.03, which is consistent with the value of KR, and implies
that the flat jet has a power law spectrum extending up to 8.4 GHz.
In the northern inner lobe (N2), we find an 
average  value $\alpha^{8.4}_{5}$ = 0.57$\pm$0.03,
and we do not distinguish between the two components found by KR.
In the southern inner lobe (S2), we separate  the flat jet with 
spectral index $\alpha^{8.4}_{5}$ = 0.74$\pm$0.03, from the surrounding region
which reaches values up to $\alpha^{8.4}_{5}$ $\sim$ 0.9.
A possibility is that the apparent jet in the southern inner lobe (S2) is a
case where the jet enters the lobe, but does not diffuse directly in it.
Rather, the jet's momentum is sufficient to carry it through the lobe, as
a distinct entity. As the jet moves around, perhaps due to tidal forces on the
nucleus, the orientation of the jet's exhaust into the lobe could change, 
permitting the lobe to be expanded from a different input angle.
Our results indicate steeper spectra than those
obtained by KR. The implication is 
that both jet components show a spectral steepening  at a frequency
beyond 5 GHz. 
It is clear from the spectral behaviour, that reacceleration processes
are much more efficient in the northern region. 
This could cause the  larger total extent of the northern
half of the source, with respect to the southern one.

\subsection {Interpretation of RM data}

The Faraday rotation of extragalactic radio sources can originate inside
the radio-emitting regions if sufficient thermal material is mixed with 
the synchrotron radiating plasma, or could be of 
external origin  if magnetic field
and thermal gas are present along the line of sight.
The interpretation of polarization data has been summarized by Laing (1984): 
if the polarization angle obeys a $\lambda^2$ law, and the rotation persists
over an angle larger than $\pi$/2, the Faraday rotation is due
to a foreground screen.


The values  of RM are generally small, ranging from 
-50 to +50 rad m$^{-2}$, with respect to the expected
foreground rotation of our Galaxy (-162 rad m$^{-2}$).
A significant contribution to the Galactic value probably
originates from the Galactic feature
crossing the region south of the galaxy associated with 3C~449 (Andernach
et al. 1992). This feature seems to affect in the same way
the rotation measure of all the region mapped by us, with no difference
between the northern and southern region.

The good $\lambda^2$ fits of the polarization angle
(Fig. 11) are in  favour of an external origin for the RM.
The foreground screen seems to be fully resolved, since there is
no significant difference between RM images at different resolutions,
and no strong depolarization is present at low frequency. 
The moderate depolarization of the jets (Sect. 3.3) could
originate within the jets, from the likely entrainment.

\begin{figure}
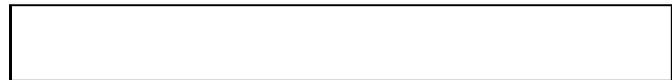

\picplace{1.0cm} 
\caption{Overlay of the grey-scale radio image at 1.365 GHz, 
with 5\arcsec~ resolution,
onto contours of the X-ray image obtained with the ROSAT PSPC. 
Contours ar drawn at 14\%, 18\%, 23\%, 30\%, 40\%, 60\%, and 80\% of the X-ray
 peak.
}
\end{figure}

A likely external Faraday screen is the intergalactic medium
associated with the galaxy group around 3C~449. 
The fact that the RM structure  is very similar in the N and S lobes is
consistent with the orientation of this radio galaxy in the plane
of the sky. 
From the analysis of X-ray data, Hardcastle et al. (1998) derived
a central gas density of 4.6 $\times$ 10$^{-3}$ cm$^{-3}$, and a core
radius of the gas distribution of 35\arcsec(18 kpc). 
Significant deviations of the RM from the Galactic value are present in the
jets up to 1\arcmin-1.1\arcmin~ from the core, i.e. about 2 core radii.
A cluster magnetic field of $\sim$0.7-0.9 $\mu$G, ordered on
scales of $\sim$5 kpc is necessary to
produce the observed values of RM.

A remarkable feature in the map of the RM is the large symmetry 
in the innermost jets, within $\sim$15\arcsec~ from the core, where
the RM assumes values more negative than the Galactic value, 
with a possible trend
that the lowest values are seen closer to the nucleus. 
From the surface brightness profile given by De Juan et al. (1994),
the galaxy brightness at a radius of about 15\arcsec~  decreases
by about a factor of 100 with respect to the peak value
(see also Fig. 2). A plausible scenario is
that the  gaseous  atmosphere associated with  the galaxy, and influencing
the jet dynamics, is responsible for the rotation measure 
and for the low fractional polarization of the innermost jets. 
To account for the observational data, we have to assume that the
magnetic field associated to the galaxy is highly tangled.

\subsection {Comparison with the X-ray emission}

In Fig. 19 the radio image of 3C~449 is overlaid to the ROSAT image,
obtained with a PSPC pointing observation, with 
an exposure of 9500 sec (Hardcastle et al.  1998). 
The data were retrieved from the public ROSAT archive.
The image was produced by binning the photon
event table in pixels of 15\arcsec, and by smoothing the map
with a gaussian 
of $\sigma$ = 45\arcsec. 
As discussed by Hardcastle et al. (1998), the large 
scale structure of 3C~449
is determined by the distribution of the hot intergalactic plasma. 
The X-ray gas is not 
spherically symmetric, and the low brightness radio emission is
anti-coincident with the X-ray filaments. The large-scale radio
emission avoids the regions of high X-ray brightness, and the radio streamers
flow where the X-ray gas is less dense. 

This effect could be due to the buoyancy of the radio plasma in
a pressure gradient. The sharp bend in 
region N4 seems to be related to the existence of a
subclump in the X-ray gas. 

\section {Conclusions}

We have presented sensitive multifrequency radio maps of the giant radio 
galaxy 3C~449, and studied its properties in total and polarized
intensity. We have mapped the source up to a total extent of 20\arcmin. 
Our conclusions are as follows.

1) The source  shows
a slightly variable unresolved core, two opposite jets, and very 
extended lobes characterized by plumes and wiggles. 
It is highly polarized at all  frequencies from 1.4 GHz to 8.4 GHz.

2) The two jets are misaligned by $\sim$11\degrees, but 
are fairly symmetrical at the resolution of 5\arcsec~
 for about 1\arcmin~ from the radio
core. They both
bend westward at $\sim$45\arcsec~ from the core, and at $\sim$ 1\arcmin~ they 
merge into the inner lobes, beyond which
 they follow a sort of helical structure.
At the highest resolution, the jets show some asymmetry,
with the southern jet being brighter on average than
the northern one. No gap of radio emission is found close to the nucleus.
The two jets are not perpendicular to the dust lane 
detected in the nucleus of the parent galaxy of 3C~449.

3) The southern and northern jets show similar collimation properties,
with a constant expansion rate at the beginning, a rapid expansion
between 6\arcsec~ and 10\arcsec~ from the core, and a recollimation
further out. The magnetic field direction
is parallel to the jet axis in the first 8\arcsec,
and then becomes transversal. 

4) From the prominence of the core power over the total radio power, we
infer that the jets are relativistic at their beginning
and the source is oriented at an angle larger than
75\degrees~ to the line of sight.
This is confirmed by the level of symmetry/asymmetry
 found in the brightness ratios of jet to counter-jet. 

5) From the application of the simple adiabatic model to the jets, evidence
of a strong jet deceleration within 10\arcsec~(5 kpc) from the nucleus
is found. A satisfactory fit to the data is found assuming an initial 
jet velocity of 0.9c, and a jet inclination to the line of sight of 
82.5\degrees. The trend of jet velocity obtained in this way is
consistent with the jet to counter-jet ratios, and with the low brightness
of the innermost jets, as due to Doppler de-boosting.
The strong jet deceleration occurs in the region where the magnetic
field changes the orientation, from longitudinal to transversal.
A likely mechanism for the jet deceleration is the entrainment of 
material from the surrounding medium.

6) The spectrum between 5 GHz and 8.4 GHz in the northern inner lobe
is flatter than in the southern inner lobe. Therefore, reacceleration processes
seem to be much more efficient in the northern than in the southern
region. This could cause the larger total extent of the northern half of the 
source, with respect to the southern one.

7) The polarization angles are well fitted 
by a $\lambda^2$ law, which persists over
about 500\degrees~ and is indication of an external origin for the RM.
Values of RM are within $\pm$50 rad m$^{-2}$ of the Galactic value. They are
likely to originate in the intergalactic medium associated
with the galaxy group of 3C~449, but also the interstellar medium of 
the galaxy itself
seems to contribute to the RM, in the innermost jet region.

\begin{acknowledgements}

We thank Dr. Greg Taylor for providing the code for the evaluation of
the rotation measure with 7 frequencies.
Thanks are due to Dr. Isabella Gioia for the optical image of the region of
the arc-like feature, and for comments on the manuscript.
We acknowledge helpful discussions with Drs. Jos\'e Luis Gomez and 
Roberto Fanti.
We thank our referee, Dr. Jacques Roland, for discussions which were
helpful in illuminating fine points of the two-fluid model.

H.A. benefitted from financial support by CONACYT (Mexico; C\'atedra
Patrimonial, ref 950093).

The National Radio Astronomy Observatory
is operated by Associated Universities, Inc., under contract with the 
National Science Foundation.
\end{acknowledgements}


\begin{thebibliography}{}


\bibitem{}
Andernach, H., Feretti, L., Giovannini, G., Klein, 
U., Rossetti, E., Schnaubelt, J., 1992, A\&AS 93, 331
\bibitem{}
 Balcells, M., Morganti, R., Oosterloo, T., P\`erez-Fournon,
I., Gonzalez-Serrano, J.I., 1995, A\&A 302, 665
\bibitem{}
Baum, S.A., O'Dea, C.P., Giovannini, G., et al., 1997, ApJ 483, 178
\bibitem{}
Begelman, M.C. 1993, in {\it Jets in Extragalactic Radio Sources}, 
H.-J. R\"oser \& K. Meisenheimer Eds., Springer-Verlag, p.145
\bibitem{}
Bicknell, G.V.: 1984, ApJ 286, 68
\bibitem{}
Bicknell, G.V., de Ruiter, H., Fanti, R., Morganti, R., Parma, P., 1990,
ApJ 354, 98
\bibitem{}
Biretta, J., 1996, in {\it Energy Transport in Radio
Galaxies and Quasars}, Hardee P.E., Bridle, A.H., Zensus, J.A. Eds.,
ASP Conference Series n. 100, p. 187
\bibitem{}
Bridle, A.H., Perley, R.A., 1984, ARA\&A 22, 319
\bibitem{}
Butcher, H.R., van Breugel, W., Miley, G.K., 1980, ApJ 235, 749
\bibitem{}
Capetti, A., Macchetto, F., Sparks, W.B., Miley, G.K.,
1994, A\&A 289, 61
\bibitem{}
Capetti, A., Celotti, A., 1998, Submitted to MNRAS
\bibitem{}
Carilli, C.L., Barthel, N., Diamond, P., 1994, AJ 108, 64
\bibitem{}
De Juan, L., Colina, L., P\'erez-Fournon, I., 1994, ApJS 91, 507
\bibitem{}
 Fanaroff, B.L., Riley, J.M.,  1974, MNRAS 167, 31P,
\bibitem{}
 Fanti, R., Lari, C., Parma, P., Bridle, A.H., Ekers, R.D., Fomalont, E.B.,
1982, A\&A 110, 169
\bibitem{}
Giovannini, G., Feretti, L., Gregorini, L., Parma, P., 1988, A\&A 199, 73
\bibitem{}
Giovannini, G., Feretti, L., Venturi, T., et al. , 1994, ApJ 435, 116
\bibitem{}
Giovannini, G., Cotton, W.D., Feretti, L., Lara, L., Venturi, T., 
Marcaide, J.M. : 1995, in {\it Quasars and Active Galactic Nuclei: High
Resolution Imaging}, M.H. Cohen \& K.I. Kellermann Eds.,
Proc. Nat. Acad. Sci. USA, Vol 92, Number 5,  p. 11356
\bibitem{}
Hardcastle, M.J., Alexander, P., Pooley, G.G., Riley, J.M, 1997,
MNRAS 288, L1 
\bibitem{}
Hardcastle, M.J., Worral, D.M, Birkinshaw, M., 1998, MNRAS 296, 1098
\bibitem{}
Hardee, P.E., Cooper, M.A., Clarke, D.A., 1994, ApJ 424, 126
\bibitem{}
 Katz-Stone, D.M., Rudnick, L., 1997, ApJ 488, 146 (KR)
\bibitem{}
Killeen, N.E.B., Bicknell, G.V., Ekers, R.D., 1986, ApJ 302, 306
\bibitem{}
Laing, R.A., 1984, in {\it Physiscs of Energy Transport in Extragalactic
Radio Sources},  A.H. Bridle \& J.A. Eilek Eds., NRAO Workshop n. 9, p. 90
\bibitem{}
 Laing R.A., 1996, in {\it Energy Transport in Radio
Galaxies and Quasars}, Hardee P.E., Bridle, A.H., Zensus, J.A. Eds.,
ASP Conference Series n. 100, p. 241
\bibitem{}
 Parma, P., De Ruiter, H.R., Fanti, R., Laing, R., 
1994, in {\it The Physics of Active Galaxies}, 
G.V. Bicknell, M.A. Dopita, P.J. Quinn Eds.,
ASP Conference Series, Vol. 54,
p. 241
\bibitem{}
Pearson, T.J., Zensus, J.A., 1987, in {\it Superluminal Radio Sources},
 J.A. Zensus \& T.J. Pearson Eds., Cambridge Univ. Press, p. 1
\bibitem{}
Pelletier, G., Roland, J., 1989, A\&A 224, 24
\bibitem{}
Pelletier, G., Roland, J., 1990, in {\it Parsec-scale Radio Jets}, Zensus 
\& Pearson Eds., Cambridge University Press, p. 323 
\bibitem{}
Pelletier, G., Sol, H., 1992, MNRAS 254, 635 
\bibitem{}
Perley, R.A., Willis, A.G., Scott, J.S., 1979, Nature 281, 437
\bibitem{}
Perley, R.A., Bridle, A.H., Willis, A.G., 1984, ApJS 54, 291
\bibitem{}
Roland, J., Lehoucq, R., Pelletier, G., 1992, in {\it Extragalactic
Radio Sources: From Beams Beams to Jets}, J. Roland, H. Sol, 
G. Pelletier Eds., Cambridge Univ. Press, p. 294
\bibitem{}
Sol, H., Pelletier, G., Asseo, E., 1989, MNRAS 237, 411
\bibitem{}
Tingay, S.J., Jauncey, D.L., Reynolds, J.E., et al., 1998, ApJ 115, 960
\bibitem{}
Urry, C.M., Padovani, P., 1995, PASP 107, 803
\bibitem{}
 Wardle, J.F.C., Kronberg, P.P., 1974, ApJ 249, 255

\end{thebibliography}
\end{document}